\documentclass[useAMS,usenatbib]{mn2e}
\usepackage{graphicx}
\usepackage{amsmath}
\usepackage{makecell}
\usepackage{subcaption}
\usepackage{macros}
\usepackage[usenames,dvipsnames,svgnames,table]{xcolor}
\usepackage{hyperref}
\definecolor{darkblue}{rgb}{0.0,0.0,0.3}
\hypersetup{colorlinks,
            linkcolor=darkblue,urlcolor=darkblue,
            anchorcolor=darkblue,citecolor=darkblue}
\usepackage{txfonts}
\DeclareSymbolFont{cmletters}{OML}{cmm}{m}{it}
\DeclareMathSymbol{v}{\mathalpha}{cmletters}{"76}
\newcommand{\RedeclareMathOperator}[2]{\rìenewcommand{#1}{}\let#1\relax\DeclareMathOperator{#1}{#2}}

\newcommand\simless\lesssim
\newcommand\simgreat\gtrsim

\newcommand\hl[1]{{#1}} 

\title[Disc Tearing and Bardeen-Petterson Alignment]{Disc Tearing and Bardeen-Petterson Alignment in GRMHD Simulations of Highly Tilted Thin Accretion Discs}

\author[Liska, Hesp, Tchekhovskoy, Ingram, van der Klis,  Markoff \& Van Moer]{M. Liska$^{1,2}$\thanks{matthewliska92@gmail.com}, C. Hesp$^{2,3,4}$, A. Tchekhovskoy$^{4}$, A. Ingram$^6$, 
  M. van der Klis$^2$, \newauthor S.B. Markoff$^2$ \&  M. Van Moer$^7$
  \\\\
$^{1}$Institute for Theory and Computation, Harvard University, 60 Garden Street, Cambridge, MA 02138, USA; John Harvard Distinguished Science and ITC Fellow\\
$^{2}$Anton Pannekoek Institute for Astronomy, University of Amsterdam, Science Park 904, 1098 XH Amsterdam, The Netherlands \\
$^{3}$Institute for Advanced Study (IAS), University of Amsterdam, Science Park 904, 1098 XH Amsterdam, The Netherlands\\
$^{4}$Amsterdam Brain and Cognition (ABC) Center, University of Amsterdam, Science Park 904, 1098 XH Amsterdam, The Netherlands\\
$^{5}$Center for Interdisciplinary Exploration \& Research in Astrophysics (CIERA),
Physics \& Astronomy, Northwestern University, Evanston, IL 60202, USA \\
$^{6}$Department of Physics, Astrophysics, University of Oxford, Denys Wilkinson Building, Keble Road, Oxford, OX1 3RH, UK\\
$^{7}$National Center for
Supercomputing Applications, University of Illinois at Urbana-Champaign,
Urbana, IL 61801, USA\\
}

\begin{document}

\date{Accepted. Received; in original form}
\pagerange{\pageref{firstpage}--\pageref{lastpage}} \pubyear{2015}
\maketitle

\label{firstpage}

\begin{abstract}
  Luminous active galactic nuclei (AGN) and X-Ray binaries (XRBs) often contain geometrically thin, radiatively cooled accretion discs. According to theory, these are -- in many cases -- initially highly misaligned with the black hole equator. In this work, we present the first general relativistic magnetohydrodynamic simulations of very thin ($h/r\sim0.015{-}0.05$) accretion discs around rapidly spinning ($a\sim0.9$) black holes and tilted by $45{-}65$ degrees. We show that the inner regions of the discs with $h/r\simless0.03$ align with the black hole equator, though out to smaller radii than predicted by analytic work. The inner aligned and outer misaligned disc regions are separated by a sharp break in tilt angle accompanied by a sharp drop in density. We find that frame-dragging by the spinning black hole overpowers the disc viscosity, which is self-consistently produced by magnetized turbulence, tearing the disc apart and forming a rapidly precessing inner sub-disc surrounded by a slowly precessing outer sub-disc. We find that the system produces a pair of relativistic jets for all initial tilt values. At small distances the black hole launched jets precess rapidly together with the inner sub-disc, whereas at large distances they partially align with the outer sub-disc and precess more slowly. If the tearing radius can be modeled accurately in future work, emission model independent measurements of black hole spin based on precession-driven quasi-periodic oscillations may become possible.

\end{abstract}


\section{Introduction}
\label{sec:introduction}
Because the gas supply of supermassive black holes (BHs) originates from far away, the angular momentum vector of the accreting gas will most likely be independent of the BH spin vector. In stellar-mass BHs, asymmetric supernova kicks can also lead to substantial misalignment between the two vectors. If the relative orientation of the two is random, the resulting accretion disc would half of the time make a 60-degree or greater angle relative to the BH equator. Tilted discs are expected in many luminous systems ranging from X-ray binaries (XRBs), active galactic nuclei (AGN), tidal disruption events (TDEs), and binary merger remnant discs \citep[e.g.][]{Hjelming1995,Orosz2001,Greene2001,Caproni2006, Caproni2007, Abbot2017}. The physics of tilted accretion systems is of crucial importance for understanding the growth of most -- if not all -- supermassive BHs throughout cosmological time, and has profound implications for jet production, BH spin measurements, and energy release of BH accretion systems (e.g. \citealt{Natarajan1998,Stella1998, Fiaconni2018}). 

\begin{figure*}
\begin{center}
\includegraphics[width=\linewidth,trim=0mm 0mm 0mm 0,clip]{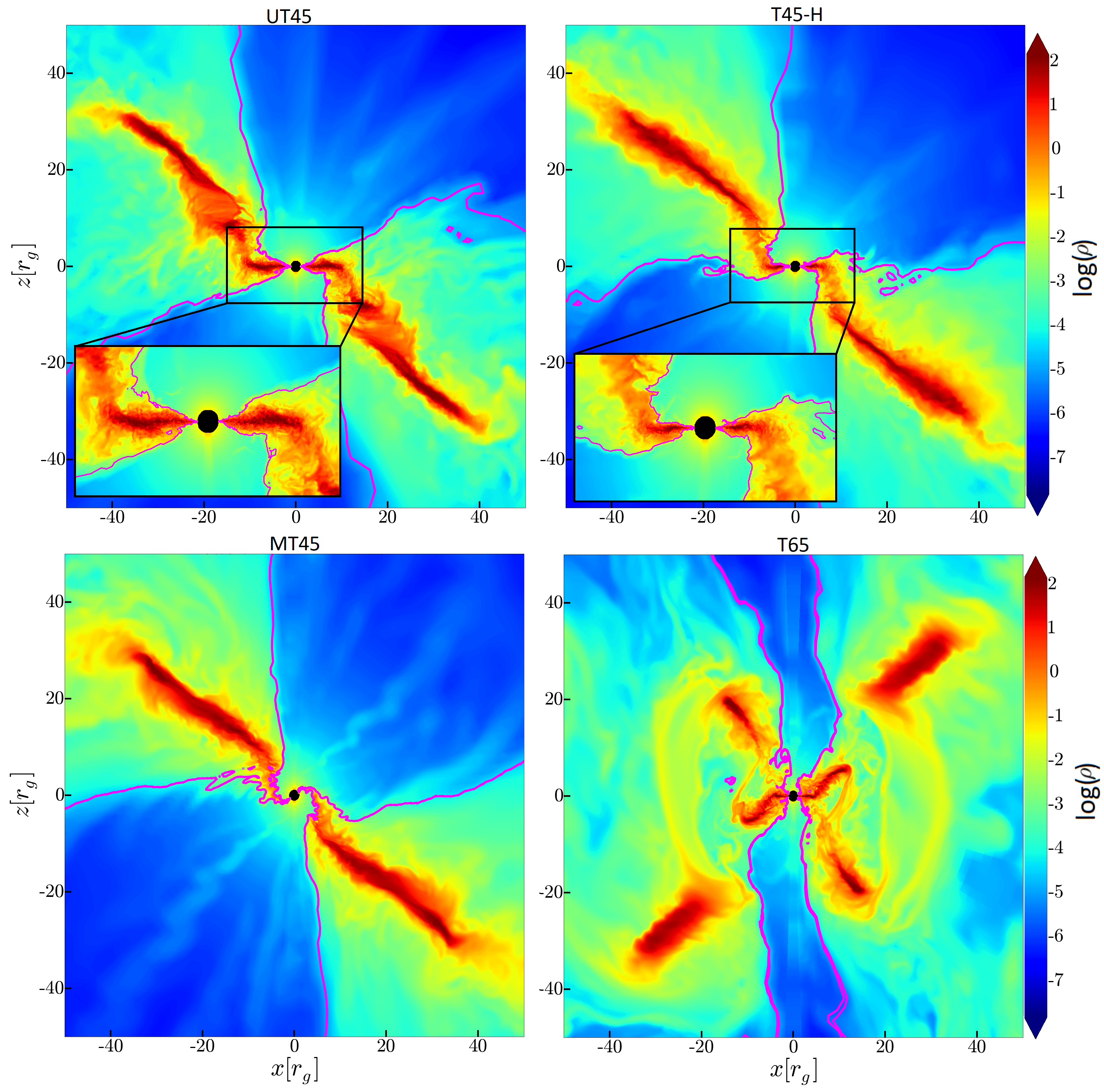}
\end{center}
\caption{Clock-wise from top-left, panels show vertical slices through fluid frame density $\log\rho$ for models UT45 and T45-H at $t=7 \times 10^4 r_g/c$, MT45 at $t=4.5 \times 10^4 r_g/c$ and T65 at $t=5.2 \times 10^4 r_g/c$ (red shows high and blue low values, see colour bar). Magenta lines indicate the jet boundary, defined as $p_b=5 \rho c^2$. In all models except the thicker, $h/r=0.05$, MT45 model, the inner disc aligns with the BH equator. The \citet{bp75} alignment radius increases with decreasing $h/r$ from $r_{\rm bp}\sim 5 r_g$ in models T45 and T45-H to $r_{\rm bp} \sim 10 r_g$ in model UT45. The transition between aligned and misaligned disc regions is very sharp, forming a `break' in tilt angle and density. In addition, the disc in model T65 tears apart into multiple, differentially precessing, sub-discs (see also Fig.~\ref{fig:tearing}). Such tears appear as discontinuities in precession angle in Fig.~\ref{fig:radplot}(k). Streamers transfer mass and angular momentum between sub-discs and directly to the BH.
}
\label{fig:contourplot}
\end{figure*}
 General relativistic frame dragging by a rotating BH warps tilted accretion discs. The evolution of such warped discs depends sensitively on the ratio between their viscosity and their dimensionless scale height, or aspect ratio \citep{Papaloizou1983}. In the analytic approach, the disc viscosity, typically parameterized through the $\alpha$-viscosity parameter, regulates the transport of angular momentum in the disc \citep{ss73}. The disc aspect ratio, $h/r$, the ratio of disc scale height $h$ to radius $r$, characterizes the internal pressure of the disc.
When the disc is relatively thick, $h/r > \alpha$, such as in the hard state of XRBs and low-luminosity AGN, the warps are transmitted by pressure waves traveling at about half the speed of sound \citep{papaloizou1995}. In this wave-like limit, analytic calculations \citep{Ivanov1997,Lubow2002} and general-relativistic magnetohydrodynamic (GRMHD) simulations \citep{Fragile2007, Texeira2014, Liska2018A, Liska2019A, White2019A} have shown that the tilt of the disc oscillates as a function of radius within about 20 gravitational radii, such that material gets accreted at high inclination angles.
When the disc is relatively thin, $h/r<\alpha$, warps are propagated through viscous diffusion, and radial tilt oscillations get damped by the dissipative effects of disc viscosity. This is thought to be the case in bright quasars and intermediate/soft state X-ray binaries.
In this limit, which is the focus of this work, a so-called Bardeen-Petterson configuration is expected to emerge: the outer disc remains tilted whereas the inner disc aligns for $\theta \lesssim \pi/2$  \citep{bp75}, or counter aligns for $\theta \gtrsim \pi/2$.  Since, by Newton's third law, the torque exerted by the BH on the inner disc is exactly equal and opposite to the torque exerted by the inner disc on the BH, \citet{bp75} alignment will not only align or counteralign the inner disc with the BH spin on shorter timescales, but will also torque the BH into alignment with the total angular momentum vector of the system on longer timescales. \hl{ In cases where the disc contains most of the angular momentum, the \citet{bp75} effect substantially accelerates the alignment between disc and BH spin axes. This can possibly lead to much faster spin-up of supermassive BHs compared to direct accretion of angular momentum through the BH's event horizon (e.g. \citealt{Natarajan1998, King2005}). However, as noted in \citet{King2005}, in cases where the disc's angular momentum is less than the black hole's angular momentum and $\theta \gtrsim \pi/2$, the disc will be torqued into counteralignment with the black hole. In such cases rapid spin-down of the central black hole is expected. This may lead to a lower than expected spin for very massive SMBHs in a chaotic accretion scenario \citep{King2006}.}

However, across both disc radius and height, there tend to be large, non-linear, and anisotropic variations of the viscous stresses induced by the magneto-rotational instability (MRI, \citealt{Balbus1991, Balbus1998}) in \emph{magnetized} accretion discs, defying the simple $\alpha-$viscosity prescription \citep[][]{Penna2010,Sorathia2010,Mckinney2012, Jiang2017}. For example, recent GRMHD simulations \citep{Liska2018C} of a very thin, $h/r=0.03$, magnetized accretion disc tilted by $10^{\circ}$ have shown that magnetic fields launch winds which counteract \citet{bp75} alignment in a strongly non-linear fashion, producing an aligned region that is much smaller than predicted for $\alpha-$discs \citep{Kumar1985, Nelson2000, Lodato2010, Nixon2012A}. 

Crucially, when a thin \emph{$\alpha-$}disc is tilted by $\mathcal{T} \gtrsim 45^{\circ}$, both simple 1D-evolution and smoothed particle hydrodynamics (SPH) simulations of warped discs suggest that frame-dragging by the spinning BH tears the disc apart into differentially precessing rings \citep{Lodato06,Nixon2012B, Nealon2015}. Just as in thin $\alpha-$discs, once a magnetized disc starts to tear, the viscosity may drop and encourage further tearing \citep{Ogilvie1999, Nixon2012A, Dogan2018}. \hl{This process has been observed in 1D calculations of warped disc evolution \citep{Lodato06}.} However, since the tilt angle exceeds the disc's angular thickness by more than an order of magnitude, the warp becomes highly non-linear -- requiring a detailed treatment involving the 3D magnetized turbulence that is the glue that holds the disc together. GRMHD simulations make this treatment possible, and we use them in this paper to study whether and how tilted, thin, magnetized discs  get torn apart. We pay special attention to the typical tearing radius and the physical prerequisites for tearing to occur. Forming such an understanding may pave the way for BH spin measurements based on quasi-periodic oscillations (QPOs) observed in XRB lightcurves \citep{Klis1989}. \hl{Such QPOs could be disc \citep{Stella1998, Ingram2009, Ingram2016, Franchini2016,Motta2017} and/or jet \citep{Kalamkar2016, Stevens2016} precession.}

In this work we present the first GRMHD simulations of highly tilted thin accretion discs in the diffusive limit of warp propagation ($h/r<\alpha$). In Section \ref{sec:numerics} we describe our code and initial conditions. We present our results in Sec.~\ref{sec:results} and conclude in Sec.~\ref{sec:conclusions}.

\section{Numerical Method and Initial Conditions}
\label{sec:numerics}
For this work we use our recently developed GRMHD code \hbox{H-AMR} \citep{Liska2018A, Chatterjee2019,hamr}. It evolves the GRMHD equations with a finite-volume--based method in modified Kerr-Schild coordinates \citep[as in][]{Gammie2003} and uses a constrained-transport scheme for magnetic field evolution \citep[see][]{Gardiner2005}. Here we employ a logarithmic spherical polar grid with $3$ to $4$ levels of adaptive mesh refinement (AMR) and $4$ levels of local adaptive time-stepping, allowing us to focus the resolution on the regions of interest. In particular, magnetized turbulence in the disc needs to be resolved \citep{Liska2018A}, so we use rest-mass density $\rho$ as the refinement criterion in order to delineate the disc \citep[as in][]{Liska2018C}. In this way, we achieve the following effective resolutions in spherical polar coordinates ($N_r \times N_{\theta} \times N_{\phi}$): $2880 \times 864 \times 1200$ in our low-resolution models, and, by doubling the resolution in every dimension,  $5760 \times 1728 \times 2400$ in our-high resolution models (see Table~\ref{table:models}). This resolves our thin discs by approximately $7$ to $14$ cells per scale height (see Sec.~\ref{sec:results} for the MRI quality factors). We use outflow boundary conditions at the inner and outer radial boundaries, which we place inside the event horizon and at $r=10^5 r_g$, respectively, where $r_g=GM/c^2$ is the gravitational radius. This way both boundaries are causally disconnected from the accretion system. Across the polar singularity in $\theta$ we use a transmissive boundary condition, which we implemented using a multi-faceted method that minimizes numerical dissipation in the polar region \citep[for details, see][]{Liska2018A}.

All models, shown in Table~\ref{table:models}, are initialized with a Kerr black hole (with $a=0.9375$) surrounded by a torus in hydrostatic equilibrium \citep[in accordance with][]{Fishbone1976} with its inner edge at $r_{\rm in}=12.5 r_g$, its pressure maximum at $r_{\rm max}=25 r_g$ (this results in torus outer edge located at $r_{\rm out}=200 r_g$), and its density normalised by setting $\max\rho=1$. 
We use the equation of state of an ideal gas, $p_{g}=(\Gamma-1)u_{g}$, where $p_{g}$ and $u_{g}$ are thermal pressure and thermal energy density, and we use a polytropic index that corresponds to a non-relativistic monoatomic ideal gas, $\Gamma = 5/3$. We seed the torus with a poloidal magnetic field defined by a covariant vector potential  $A_{\phi}=(\rho-0.05)^{2}r^{3}$. We normalize the magnetic field strength by requiring that $\max p_{g}/\max p_{b}=30$, where $p_{b}$ is the magnetic pressure. Subsequently, we tilt the torus and magnetic field relative to the BH spin (and the grid) by an angle $\mathcal{T}_{\rm init}$ (see \citealt{Liska2018A} for details). Finally, we reduce the disc thickness to a target scale-height $h/r$ by cooling the gas at a rate slow enough to avoid disruption of the disc orbital dynamics. \hl{We do this by letting the internal energy decay throughout the simulation exponentially over time, with the time constant set by the orbital timescale \citep{Noble2009}. 
Since the disc needs some time to cool and reach the target scale height, we only include data after $t=10^4 t_g$ into our analysis, where $t_g = r_g/c$. } We initialize the high-resolution models \hbox{T45-H} and UT45 with a well-evolved state of model T45 at $t = 4.8 \times 10^4 t_g$ (see Table~\ref{table:models} for details). In the case of T45-H, we, additionally, reduce the target thickness in the cooling function from $h/r=0.03$ to $h/r=0.015$. 

\begin{table}
\centering
\label{Physical parameters}
\begin{tabular}{llllll}
\hline
Model & Full name & $\mathcal{T}_{\rm init}$ &  $N_r \times N_{\theta} \times N_{\phi}$ & $h/r$ & $t_{i}$-$t_{f}$ [$10^4 t_g$]\\
\hline
T45& T45HR03L& $45^{\circ}$ & 2880$\times$864$\times$1200 & $0.03$  & $0$-$10.5$\\
T45-H& T45HR03H & $45^{\circ}$ & 5760$\times$1728$\times$2400 & $0.03$ & $4.8$-$7.2$\\
UT45& T45HR015H & $45^{\circ}$ & 5760$\times$1728$\times$2400 & $0.015$ & $4.8$-$7.0$\\
MT45& T45HR05L & $45^{\circ}$ &  2880$\times$864$\times$1200 & $0.05$ & $0$-$4.5$\\
T65& T65HR03L & $65^{\circ}$ & 2880$\times$864$\times$1200 & $0.03$ & $0$-$12$\\

\hline
\end{tabular}
\caption{
The tilt ($\mathcal{T}_{\rm init}$), number of cells in $r-$, $\theta-$ and $\phi-$ coordinates ($N_{r} \times N_{\theta} \times N_{\phi}$), disc thickness ($h/r$) and time interval ($t_i-t_f$) for each model. 
}
\label{table:models}
\end{table}

\begin{figure}
\begin{center}
\includegraphics[width=\linewidth,trim=0mm 0mm 0mm 0,clip]{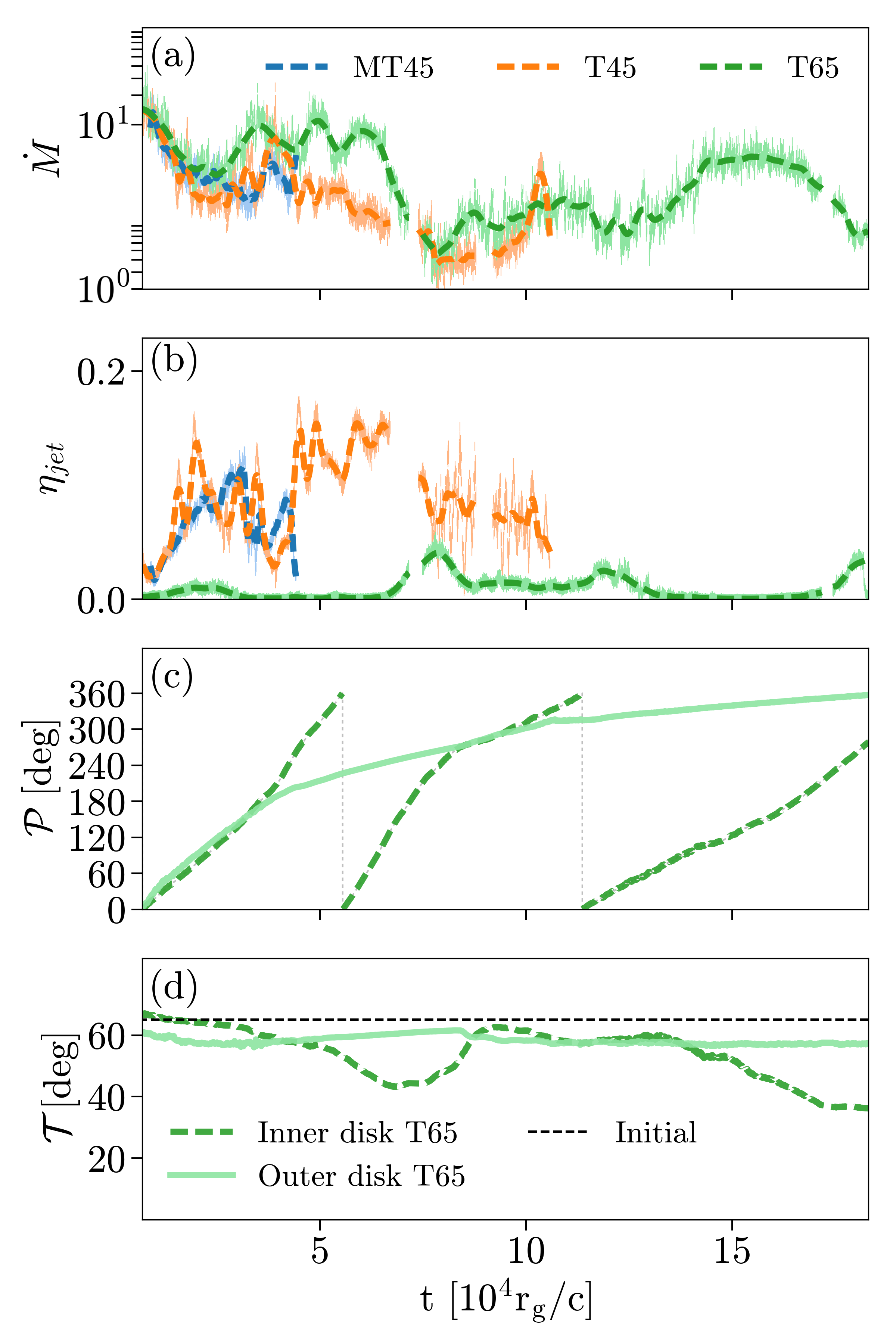}
\end{center}
\caption{Time evolution for models T65 (green), T45 (orange) and MT45 (blue). [panel (a)] The BH mass accretion ($\dot{M}_{\rm BH}$) is about $1.5$-$3$ times higher in T65 compared to the other two models, presumably due to cancellation of angular momentum when the sub-discs become partially opposed (Fig.~\ref{fig:tearing}). [panel (b)] The jet efficiency ($\eta_{\rm jet}$, measured at $r \sim 10 r_g$) is reduced significantly in T65 due to dissipation when the jet gets reoriented at the tearing radius of $r \sim 5-30 r_g$. [panel (c)] The precession angle ($\mathcal{P}$) in model T65 of the two outermost sub-discs seen in Fig.~\ref{fig:tearing}
increases much more rapidly than the precession angle of the outer sub-disc (solid) due to the differential nature of the \citet{LT1918} torque. [panel (d)] The inner sub-disc quickly starts aligning with the BH spin (i.e., decreasing $\mathcal{T}$) during phases of disc tearing – when there is a large difference in $\mathcal{P}$ between the inner and outer sub-discs (e.g., between 5-7.5 $\times 10^4 t_g$). After the inner sub-disc makes a full cycle in $\mathcal{P}$, it conjoins again with the outer sub-disc and is torqued back into a tilted configuration.
}
\label{fig:timeplot}
\end{figure}

\begin{figure*}
\begin{center}
\includegraphics[width=\linewidth,trim=0 0.5cm 0cm 0cm,clip]{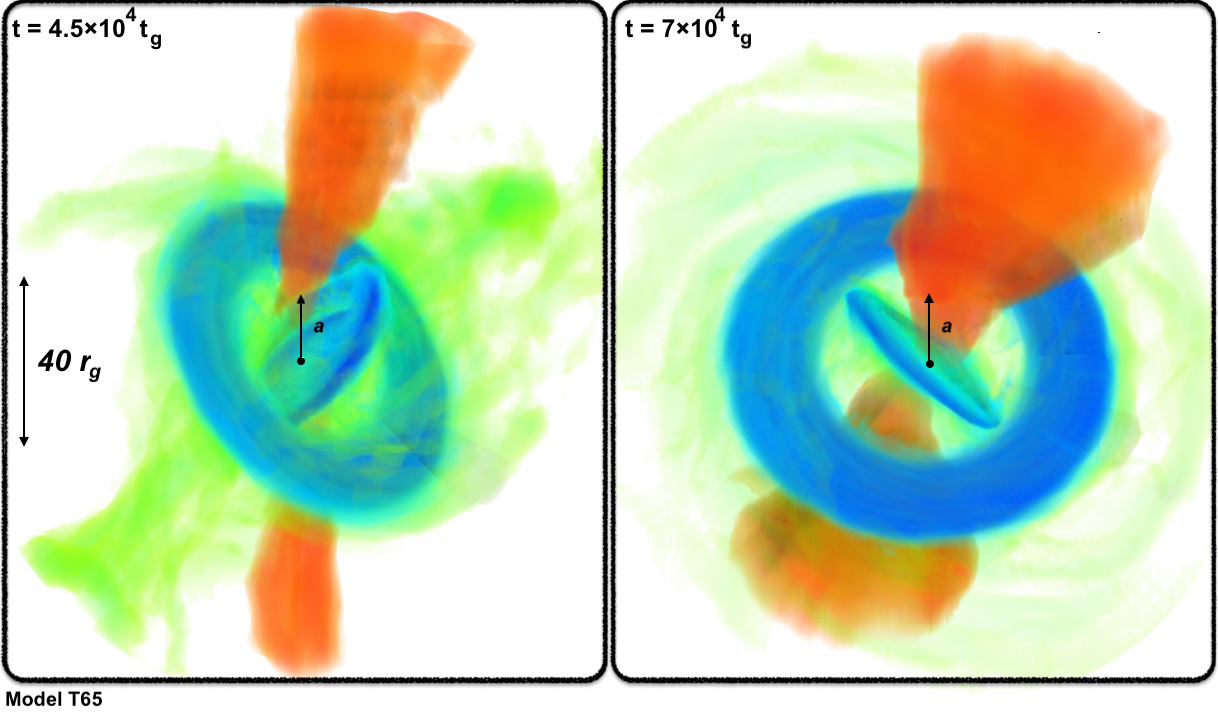}
\end{center}
\caption{First demonstration that a highly tilted magnetized accretion disc (blue) can tear up into multiple, radially extended, sub-discs. In this volume rendering of model T65 at $t=4.5 \times 10^4 r_g/c$ (left) and $t=7 \times 10^4 r_g/c$ (right) the BH spin $\boldsymbol a$ points up (black arrows) and the length scale of 40 $r_g$ is indicated on the left. The jets (red) are launched along the direction of the inner sub-disc, but, as they propagate outwards, tend to align with the corona (green), which is aligned with the outer sub-disc. While changing orientation, the jets exert an equal and opposite force on the outer sub-disc, pushing it onto more energetic orbits.}
\label{fig:tearing}
\end{figure*}

\begin{figure*}
\begin{center}
\includegraphics[width=1\linewidth,trim=0mm 0mm 0mm 0,clip]{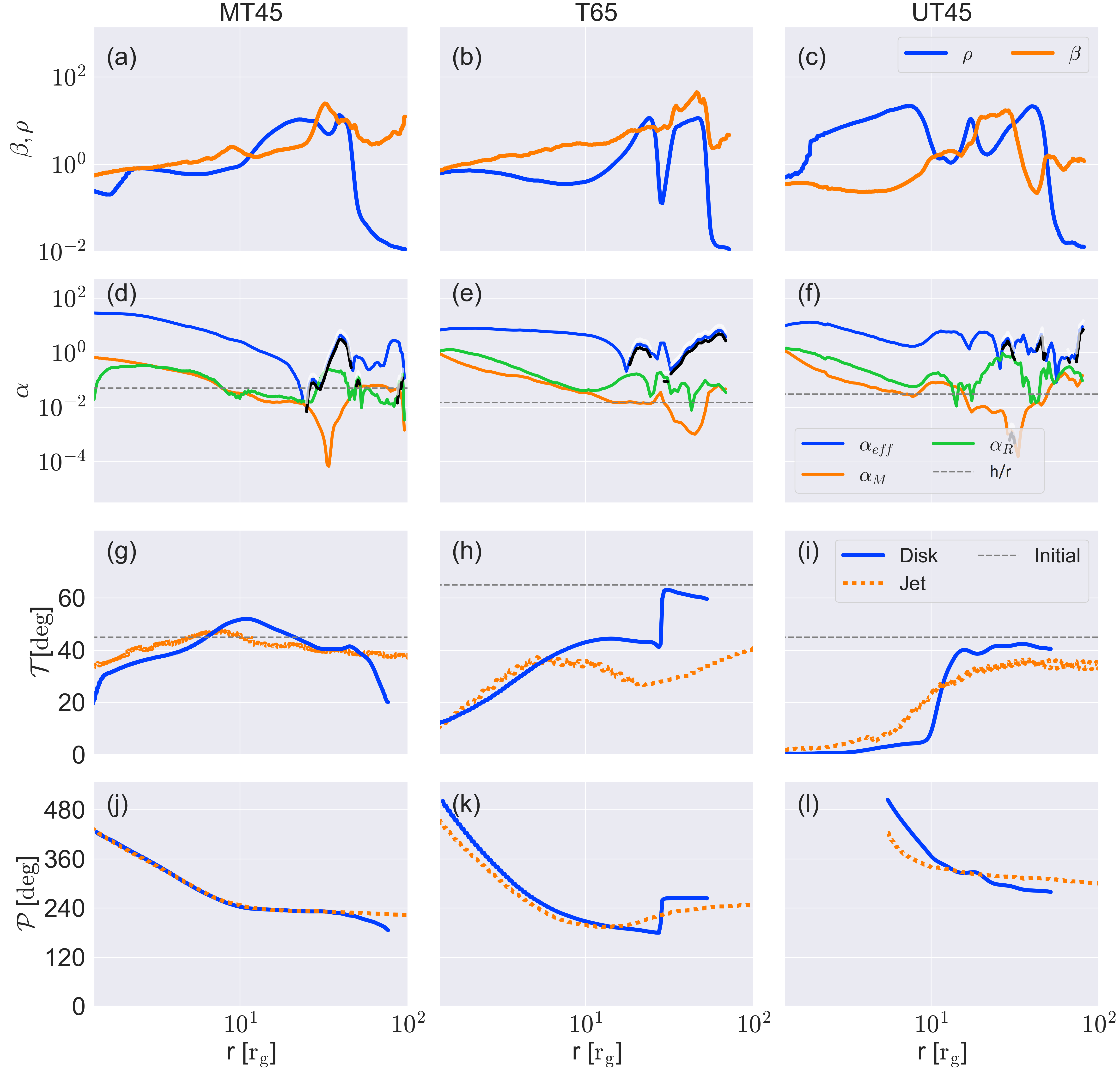}
\end{center}
\caption{Radial profiles for models MT45 (left), T65 (middle), and UT45 (right) averaged over $4.2-4.4 \times 10^4 r_g/c$ (MT45) and $6.8-7.0 \times 10^4 r_g/c$ (T65 and UT45).
[(a-c)] In all models the disc remains mostly gas-pressure dominated ($\beta \gtrsim 1$) allowing it to achieve the target scale-height. The density ($\rho$) drops around the disc breaking/tearing radius. 
[(d-f)] The effective viscosity ($\alpha_{\rm eff}$) exceeds the sum of the Maxwell ($\alpha_M$) and Reynolds ($\alpha_R$) stresses, presumably due to the presence of wind driven torques and, in T65, cancellation of angular momentum when misaligned angular momentum flow between sub-discs. Negative values of the effective viscosity ($\alpha_{\rm eff} \propto v_r$; black-white emphasis on the plots) are obtained when a particular annulus of the disc has net local outward movement (due to angular momentum transport). As a reference we also plot the target scale height of the disc ($h/r$).
[(g-l)] The disc's and jet's tilt ($\mathcal{T}$) and precession ($\mathcal{P}$) angles tend to follow the same trend, since the outer disc torques the jet into (partial) alignment with itself. The discontinuity in tilt angle between inner and outer sub-disc in T65 occurs due to the angular momentum cancellation when the inner and outer sub-disc are partially opposed due to differential precession. It is unrelated to \citet{bp75} alignment.
}
\label{fig:radplot}
\end{figure*}
\section{Results}
\label{sec:results}
Figure~\ref{fig:contourplot} shows a colour map of the density for models UT45, \hbox{T45-H}, MT45 and T65 in their evolved state at $t\gtrsim4 \times 10^4 r_g/c$. As can be seen from 3D animations (see \href{https://www.youtube.com/playlist?list=PLDO1oeU33GwlaPSME1TdCto1Y3P6yG91L}{this YouTube playlist}), the inner disc always aligns with the BH spin in models T45-H and UT45, aligns most of the time in model T45, some of the time in model T65, and does not align at all in model MT45. This can be understood because the disc is thicker in model MT45, $h/r=0.05$, and may fall outside the diffusive warp propagation regime, $h/r < \alpha$, where the \citet{bp75} alignment is expected (e.g. \citealt{Ivanov1997}).

For our thinnest discs, we can establish the dependence of disc structure on thickness. In particular, we observe that the \citet{bp75} alignment radius, $r_{\rm bp}$, increases from $r_{\rm bp} \sim 5 r_g$ at $h/r=0.03$ (models T45 and T45-H) to $r_{\rm bp} \sim 10 r_g$ at $h/r=0.015$ (model UT45). While this is consistent with the predicted analytic scaling of $r_{\rm bp}\sim (h/r)^{-8/7}$  (e.g. \citealt{Kumar1985}), the proportionality constant for our simulation is much smaller than in the analytic scaling. This discrepancy might be due to the torques applied on the disc by the large-scale magnetically-powered disc outflows that can counteract the \citet{bp75} alignment and reduce the value of $r_{\rm bp}$ \citep{Liska2018C}.

We note that there are important qualitative differences in the simulation results presented here and our previous work, which considered a disc with exactly the same initial conditions but tilted by a much smaller angle, $\mathcal T = 10^{\circ}$ \citep{Liska2018C}. For this smaller value of tilt, the inner aligned and outer misaligned parts of the accretion disc were separated by a smooth warp \citep{Liska2018C}. However, for much larger tilt angles considered here, $\mathcal T=45{-}65$ degrees, the inner aligned and outer misaligned parts of the disc are separated by a discontinuity in tilt angle and sharp drop in density. This transition is referred to as a \emph{disc break} (e.g., \citealt{Lodato2010,Nixon2012A}). As we discuss in Sec.~\ref{sec:conclusions}, the development of a break will likely change the physical properties and observational signatures of a tilted disc and its precession. 

The \citet{bp75} alignment is variable in time: as seen from the movies, as a general trend, an increase/decrease in BH mass accretion rate $\dot{M}_{\rm BH}$ (Fig.~\ref{fig:timeplotplot}a) tends to accompany the periods of misalignment/alignment of the inner disc with the BH in models T45 and T65. This might be due to rapid accretion giving the disc insufficient time to align with the BH spin axis (as postulated in e.g. \citealt{Nealon2015, Liska2018C}). Alternatively, the formation of a break may suppress accretion making the drop in mass accretion rate a consequence of the alignment instead of a cause.

Figures~\ref{fig:contourplot} and \ref{fig:tearing} show that at large values of tilt, $\mathcal T = 65^{\circ}$, in model T65, the disc tears into $2$ or sometimes even $3$ differentially precessing sub-discs at radii ranging from $5r_g$ to $30 r_g$. This is caused by the differential \citet{LT1918} torques exceeding the viscous torques that hold the disc together. We observe that the disc also tears at lower tilt values, $\mathcal T =45^{\circ}$, in models T45 and UT45: however, this happens at later times, $t \gtrsim 10^5 r_g/c$ for T45 and $t \gtrsim 7 \times 10^4 r_g/c$ for UT45, respectively. To reach the BH, the gas must pass between sub-discs via streamers, which are tentacle-like low density structures connecting adjacent sub-discs, as seen in Figs~\ref{fig:contourplot} and \ref{fig:tearing}. Note that the BP-aligned part of the inner sub-disc in  Fig~\ref{fig:contourplot} for T65 is fed partially by the misaligned outer part of the inner sub-disc and partially directly by the middle sub-disc, bypassing the misaligned part of the inner sub-disc. We leave quantification of mass transfer rate between the different sub-discs to future work.

As the phase difference in precession angle $\mathcal{P}$ between the inner and outer sub-disc(s) builds up (Fig.~\ref{fig:timeplotplot}c), this contributes to effective cancellation of angular momentum between the two (\citealt{Nixon2012counter,Nixon2012B}). This might explain a factor of $1.5{-}3$ enhancements in BH mass accretion rate (Fig.~\ref{fig:timeplotplot}a) at $t \sim 5 \times 10^4 r_g/c$ and $t \sim 15 \times 10^4 r_g/c$ in model T65, corresponding to the moments in time when $\mathcal P_{\rm inner}-\mathcal P_{\rm outer} = 180 + 360n$, $n = 0,1$ degrees, i.e., when the sub-discs are counter-rotating. Interestingly, the evolution of a sub-disc of radius $10 r_g$ formed at $t \sim 4.5 \times 10^4 r_g/c$ (see the left panel of Fig.~\ref{fig:tearing}) differs from the medium-sized sub-disc of radius $20-30 r_g$ present throughout the simulation. While the smaller sub-disc slowly accretes into the BH, the medium-sized sub-disc temporarily merges with the outer sub-disc around $t \sim 8 \times 10^4 r_g/c$ for a duration of $t \sim 3.0 \times 10^4 r_g/c$ (both in precession angle $\mathcal{P}$ and tilt angle $\mathcal{T}$, see Fig.~\ref{fig:timeplotplot}c,d) before tearing again, but this time at a slightly larger radius of $r \sim 30 r_g$. This merging may be driven by density fluctuations in the inner disc. Namely, during each precession cycle a significant portion of the inner disc mass falls into the BH (viscous timescale of inner disc equals $\sim 1-3 \times 10^4 r_g/c$), which reduces the differential \citet{LT1918} precession rate between inner and outer disc, allowing them to merge and subsequently tear at a different radius.

To gain a better insight into the internal disc dynamics in models MT45, UT45 and T65 we analyze radial profiles of density, plasma $\beta=p_g/p_B$ and $\alpha-$viscosity, as shown in Fig.~\ref{fig:radplot}(a)-(d). Here, all vectors are calculated in a coordinate system ($r$, $\tilde{\theta}$, $\tilde{\phi}$) aligned with the local rotation axis of the disc (see \citealt{Liska2018C} for details). Since $\beta \gtrsim 1$ for $r>10 r_g$ the disc remains (mostly) gas pressure dominated, except where the disc breaks or tears and the density drops. As expected for gas pressure dominated discs, we verified that the density weighted scale height, $(h/r)_{\rho}=\langle\tilde{\theta}-\langle\tilde{\theta}\rangle_{\rho} \rangle_{\rho}$, matches the thermal scale height, $(h/r)_{\rm thermal}=\langle c_{s} \rangle_{\rho}/\langle v_{k}\rangle_{\rho}$, where $v_k$ is the Keplerian 3-velocity, $c_s$ is the sound speed, and $\langle\dots\rangle_q$ indicates an angle-average weighted by the quantity $q$. Throughout the disc in models UT45 and T65 the sum of the Maxwell, $\alpha_M=b^r b^{\tilde{\phi}}/(p_g+p_b)$, and Reynolds, $\alpha_R=\rho u^r u^{\tilde{\phi}}/(p_g+p_b)$, stress contributions to the viscosity parameter remains larger than the disc scale height. This confirms that these two discs are in the $h/r<\alpha$ diffusive warp propagation regime \citep{Papaloizou1983}. Here, $b^\mu$ and $u^\mu$ are the magnetic and velocity 4-vectors.

Surprisingly, the effective viscosity parameter, $\alpha_{\rm eff}=-v_r v_k/ c_{s}^2$, which is a measure of the radial inflow speed, exceeds the sum of $\alpha_R$ and $\alpha_M$ by more than an order of magnitude. This indicates that the angular momentum transport in turbulent discs cannot be described by stresses induced through a local $\alpha-$viscosity. Most likely, large-scale magnetic and/or internal torques contribute to this discrepancy by transporting angular momentum outwards. The situation is similar to a model of a weakly tilted disk of \citet[this model started out with a tilt angle of $\mathcal{T} = 10^\circ$, but at late times the tilted reduced to $\mathcal T \sim 2^\circ$]{Liska2018C}, but with several times smaller  discrepancy between the effective viscosity parameter and combined Reynolds/Maxwell stresses 
than in this work. This suggests that tilt-related effects, such as viscous dissipation in warps (e.g. \citealt{Lodato06}) and/or spiral shocks aligned with the line of nodes \citep{Fragile2008, White2019A}, are likely the more dominant dissipation mechanism at large tilt angles considered in this work ($\mathcal T\gtrsim45^\circ$). Additionally, in our model T65, cancellation of angular momentum where the sub-discs intersect could explain the enhanced accretion rate during episodes of large misalignment between adjacent sub-discs (see Fig.~\ref{fig:timeplotplot}). Observationally, the large, above unity, effective viscosity in the inner disc may cause accretion to proceed so fast that the electrons and ions do not have the time to equilibrate their temperatures, leading to a two-temperature plasma (\citealt{Esin1997}, see also Sec.~\ref{sec:conclusions}), forming a rapidly precessing advection-dominated accretion flow that produces a hard spectrum \citep{Narayan1994}.

To verify numerical convergence, we compared the radial profiles of our model T45 and model T45-H, carried out a twice as high resolution, and found a good level of agreement. In addition, the number of cells per MRI wavelength, $Q_r \times Q_{\tilde{\theta}} \times Q_{\tilde{\phi}}$, saturates around $100\times100\times500$ for $r\simless20r_g$ and $10\times10\times150$ for $r\simgreat20r_g$ in our low resolution models, indicating that the MRI turbulence is well-resolved  in the inner and reasonably well-resolved in the outer regions of the accretion disc \citep{Sorathia2010, Hawley2011}.

All our models launch moderately strong \citet{bz77} jets, which we define as magnetically-dominated regions with $p_b/ \rho c^2 >5$. Figure~\ref{fig:timeplotplot}(b) shows that their energy outflow efficiency, or power measured in units of accretion power $\dot{M}_{\rm BH} c^2$, reaches $\eta_{\rm jet} \sim 1\%-10\%$.
Interestingly, the jets appear to follow the orientation of the disc over a wide range of length scales. For instance, jets at small radii align with the inner disc, as seen in Figs~\ref{fig:tearing} and \ref{fig:radplot}(g-i). At larger radii, they get torqued via the corona, (roughly) defined as all gas not part of the jet and whose density is a factor $10^3$ smaller than that in the disc  (see also \citealt{Liska2019A}), into (partial) alignment with the outer disc. This is also seen in Fig.~\ref{fig:tearing}. In fact, in model T65, the disc-jet interaction can be so strong that the jets running into outer sub-disc can push it into higher orbits {(see \href{https://www.youtube.com/playlist?list=PLDO1oeU33GwlaPSME1TdCto1Y3P6yG91L}{this YouTube playlist})}. This may deprive the BH of its mass supply and quench the accretion at times beyond those simulated.

In nature, the outermost sub-disc would be essentially infinite in size, definitely much larger than in our simulations, and would have an essentially infinite precession period \citep{Liska2018A}. How does this affect potential quasi-periodic signals coming from such systems? Due to the lack of precession of the extremely large outermost sub-disc, all precessing sub-discs would be located at smaller radii. Thus, only the parts of the jet outside of the aligned region, $r\gtrsim r_{\rm bp}$, and inside of the outermost, non-precessing sub-disc would be expected to contribute to jet-driven QPOs (e.g., \citealt{Kalamkar2016,Stevens2016}). 

\section{Discussion and conclusion}

\label{sec:conclusions}
In this work we have presented the first GRMHD simulations of highly tilted ($\mathcal T = 45{-}65$~degrees), thin ($h/r = 0.015 {-} 0.05$) accretion discs around rapidly spinning BHs ($a=0.9375$). We demonstrate for the first time that in the presence of realistic magnetized turbulence the inner parts of such discs can align with the BH spin axis, as predicted by \citet{bp75}. However, we find that the alignment radius of $r_{\rm bp} \lesssim 5{-}10 r_g$ (Fig. \ref{fig:contourplot}) is much smaller than predicted by analytic models. This discrepancy may be caused by a very large effective viscosity $\alpha_{\rm eff}$ giving the inner disc insufficient time to align (see \citealt{Liska2018C} for discussion). 

The \citet{bp75} effect can torque the BH into alignment with the outer disc \citep[as long as the disc's angular momentum exceeds the BH's angular momentum, see][]{Natarajan1998,King2005, Fiaconni2018} and lead to rapid BH spinup in such systems. Rapidly spinning black holes are known to launch powerful \citet{bz77} jets in the presence of large scale poloidal or toroidal magnetic flux (e.g., \citealt{Tchekhovskoy2011,Mckinney2012,Liska2018B}). Consistent with semi-analytic work \citep{Nixon2012A} and smoothed-particle hydrodynamics (SPH) simulations \citep{Lodato2010} the transition between the inner aligned and outer misaligned discs occurs over a very short distance and exhibits a sharp drop in density. As gas crosses this `break', misaligned angular momentum cancels rapidly, possibly leading to enhanced dissipation of kinetic and magnetic energy causing non-thermal emission. \hl{We note that the absence of a break in recent MHD simulations of a thin ($h/r\sim0.05$) disk tilted by $\mathcal{T}\sim24^{\circ}$ \citep{Hawley2019} is not inconsistent with our work, as evidenced by the absence of a break in model MT45.}

When the disc tilt is large, $\mathcal{T}\gtrsim45^{\circ}$, we show for the first time that a \emph{magnetized} thin $h/r=0.015{-}0.03$ disc can tear into multiple independently precessing sub-discs, as seen in Fig.~\ref{fig:tearing}. In future work we will investigate whether tearing can happen at smaller values of disc tilt for thinner discs and whether, therefore, even weakly misaligned discs in XRBs and AGN can be subject to disc tearing.
Observationally, we expect disc tearing to lead to a wide range of interesting phenomena. Differential precession, as explained in Sec.~\ref{sec:results}, can lead to cancellation of angular momentum leading to a factor of few increase in the mass accretion rate. This may explain flaring in the hard-intermediate/ultra-luminous state of XRBs \citep{Remillard2006,McClintock2006}. For instance if a disc undergoes several tearing events in quick succession, its luminosity will increase. However, the  inner disc density eventually drops, because the outer disc is unable able to keep up the supply of gas indefinitely, especially when a powerful jet injects energy and angular momentum into the outer disc (Sec.~\ref{sec:results}). This drop in density, together with a very short accretion time (due to unusually large effective viscosity, $\alpha_{\rm eff}\gtrsim1$, in Fig.~\ref{fig:radplot}c,d), may lead to the decoupling of ions and electrons into a two-temperature plasma, reducing the ability of ions in the disc to cool and puffing up the inner thin disc into a hotter, and less radiatively efficient, thick advection-dominated accretion flow (ADAF, see \citealt{Narayan1994}). In fact, for $\alpha_{\rm eff}>1$ a thick disc would be expected to always form when two-temperature thermodynamics effects are taken into account (see e.g. \citealt{Esin1997,Fereira2006, Marcel2018A, Marcel2018B, Liska2018C}). Since the viscous torque is stronger for a larger disc thickness, the thicker disc would no longer undergo tearing. \hl{For instance, in agreement with \citet{Hawley2019}, our thicker disc model MT45 with $h/r=0.05$, seen in the bottom-left panel of Fig.~\ref{fig:contourplot}, does not show any signs of tearing.} In the absence of disc tearing, the connection with the outer thin disc gets reestablished. The outer disc then feeds the inner one, and the BH mass accretion rate rises. This results in the density increase of the inner disc and its rapid cooling and collapse into a thin disc. The cycle then repeats. During each such cycle magnetic jets violently interact with the precessing sub-discs, making the tearing radius an interesting location for enhanced dissipation and (non-)thermal emission. Additionally, the streamers connecting torn sub-discs to each other can scatter and/or reradiate the emission from the central regions, substantially affecting the emergent spectrum and variability and making the discs appear larger than otherwise. This may resolve the puzzle of what makes AGN disc sizes exceed the predictions of an $\alpha-$disc model \citep[e.g.,][]{2011ApJ...729...34B}. 

\hl{It has been suggested that BH spin could be measured based on precession induced Type-C QPOs \citep{Stella1998,Ingram2009,Franchini2016, Motta2017}, whose frequency depends on BH spin and disc size.} Such measurements would be able to independently verify the accuracy of the continuum fitting (e.g. \citealt{Mcclintock2013}) and iron-line methods (e.g. \citealt{Reynolds2008}) without making any assumptions about the disc's emission near the innermost stable circular orbit (ISCO) or assuming that the system is aligned. Making self-consistent predictions for the tearing radius, based on e.g. disc thickness, tilt and magnetic field topology, requires clear theoretical understanding of the physics driving disc tearing. On a basic level, for a disc to tear, the differential \citet{LT1918} torques need to exceed the viscous torques holding the disc together. For $\alpha$-discs, the effective torque counteracting breaking and tearing can be derived as function of warp amplitude \citep{Ogilvie1999} making it possible to calculate criteria for disc breaking and tearing \citep{Dogan2018}. However, disc tearing in GRMHD leads to a substantially different morphology compared to SPH simulations.
More specifically, instead of tearing up into narrow `rings' with $\Delta r \sim h$ as seen in SPH models in the $h/r < \alpha$ regime \citep{Nixon2012B}, our GRMHD models form radially extended sub-discs with $\Delta r \gg h$ (Fig.~\ref{fig:tearing}). We note that while SPH simulations in the thick disk, $h/r > \alpha$, regime form rings which are more radially extended \citep{Nealon2015}, their radial extent is still much smaller than presented in this work. This discrepancy suggests the disc tearing process may be more complicated when magnetized turbulence self-consistently determines the viscous-like coupling within the disc. We hypothesize that this might be caused by radial tension along magnetic field lines, which is neglected in the $\alpha$-disc approximation. These and other questions, such as how radiation pressure influences the behaviour of disc tearing into sub-discs, how a disc break influences the angular momentum transport, and what causes the apparent stochasticity in the disc tearing radius (see Sec.~\ref{sec:results}), will be addressed in future work.

\section{Acknowledgments}
This research was made possible by NSF PRAC awards no.~1615281 and OAC-1811605 as part of the Blue Waters sustained-petascale computing project, which is supported by the National Science Foundation (awards OCI-0725070 and ACI-1238993) and the state of Illinois. Blue Waters is a joint effort of the University of Illinois at Urbana-Champaign and its National Center for Supercomputing Applications.
ML and MK were supported by the NWO Spinoza Prize, AI by the Royal Society URF, CH by the NWO Research Talent grant (no. 406.18.535), SM by the NWO VICI grant (no. 639.043.513), and AT by the NSF grants 1815304, 1911080 and NASA grant
80NSSC18K0565. The simulation data presented in this work is available upon request to AT at \href{mailto:atchekho@northwestern.edu}{atchekho@northwestern.edu}.

\section{Supporting Information}
Additional Supporting Information may be found in the online version
of this article: movie files.
See our \href{https://www.youtube.com/playlist?list=PLDO1oeU33GwlaPSME1TdCto1Y3P6yG91L}{YouTube
  playlist} for 3D visualizations of all models.

\bibliography{mybib,sasha}

\begin{thebibliography}{}
\makeatletter
\relax
\def\mn@urlcharsother{\let\do\@makeother \do\$\do\&\do\#\do\^\do\_\do\%\do\~}
\def\mn@doi{\begingroup\mn@urlcharsother \@ifnextchar [ {\mn@doi@}
  {\mn@doi@[]}}
\def\mn@doi@[#1]#2{\def\@tempa{#1}\ifx\@tempa\@empty \href
  {http://dx.doi.org/#2} {doi:#2}\else \href {http://dx.doi.org/#2} {#1}\fi
  \endgroup}
\def\mn@eprint#1#2{\mn@eprint@#1:#2::\@nil}
\def\mn@eprint@arXiv#1{\href {http://arxiv.org/abs/#1} {{\tt arXiv:#1}}}
\def\mn@eprint@dblp#1{\href {http://dblp.uni-trier.de/rec/bibtex/#1.xml}
  {dblp:#1}}
\def\mn@eprint@#1:#2:#3:#4\@nil{\def\@tempa {#1}\def\@tempb {#2}\def\@tempc
  {#3}\ifx \@tempc \@empty \let \@tempc \@tempb \let \@tempb \@tempa \fi \ifx
  \@tempb \@empty \def\@tempb {arXiv}\fi \@ifundefined
  {mn@eprint@\@tempb}{\@tempb:\@tempc}{\expandafter \expandafter \csname
  mn@eprint@\@tempb\endcsname \expandafter{\@tempc}}}

\bibitem[\protect\citeauthoryear{{Abbott} et~al.,}{{Abbott}
  et~al.}{2017}]{Abbot2017}
{Abbott} B.~P.,  et~al., 2017, \mn@doi [Physical Review Letters]
  {10.1103/PhysRevLett.118.221101}, \href
  {http://adsabs.harvard.edu/abs/2017PhRvL.118v1101A} {118, 221101}

\bibitem[\protect\citeauthoryear{{Balbus} \& {Hawley}}{{Balbus} \&
  {Hawley}}{1991}]{Balbus1991}
{Balbus} S.~A.,  {Hawley} J.~F.,  1991, \mn@doi [\apj] {10.1086/170270}, \href
  {http://adsabs.harvard.edu/abs/1991ApJ...376..214B} {376, 214}

\bibitem[\protect\citeauthoryear{Balbus \& Hawley}{Balbus \&
  Hawley}{1998}]{Balbus1998}
Balbus S.~A.,  Hawley J.~F.,  1998, \mn@doi [Rev. Mod. Phys.]
  {10.1103/RevModPhys.70.1}, 70, 1

\bibitem[\protect\citeauthoryear{{Bardeen} \& {Petterson}}{{Bardeen} \&
  {Petterson}}{1975}]{bp75}
{Bardeen} J.~M.,  {Petterson} J.~A.,  1975, \mn@doi [\apjl] {10.1086/181711},
  \href {http://adsabs.harvard.edu/abs/1975ApJ...195L..65B} {195, L65}

\bibitem[\protect\citeauthoryear{{Blackburne}, {Pooley}, {Rappaport}  \&
  {Schechter}}{{Blackburne} et~al.}{2011}]{2011ApJ...729...34B}
{Blackburne} J.~A.,  {Pooley} D.,  {Rappaport} S.,   {Schechter} P.~L.,  2011,
  \mn@doi [\apj] {10.1088/0004-637X/729/1/34}, \href
  {http://adsabs.harvard.edu/abs/2011ApJ...729...34B} {729, 34}

\bibitem[\protect\citeauthoryear{{Blandford} \& {Znajek}}{{Blandford} \&
  {Znajek}}{1977}]{bz77}
{Blandford} R.~D.,  {Znajek} R.~L.,  1977, \mn@doi [\mnras]
  {10.1093/mnras/179.3.433}, \href
  {http://adsabs.harvard.edu/abs/1977MNRAS.179..433B} {179, 433}

\bibitem[\protect\citeauthoryear{{Caproni}, {Abraham}  \& {Mosquera
  Cuesta}}{{Caproni} et~al.}{2006}]{Caproni2006}
{Caproni} A.,  {Abraham} Z.,   {Mosquera Cuesta} H.~J.,  2006, \mn@doi [\apj]
  {10.1086/498684}, \href {http://adsabs.harvard.edu/abs/2006ApJ...638..120C}
  {638, 120}

\bibitem[\protect\citeauthoryear{{Caproni}, {Abraham}, {Livio}  \& {Mosquera
  Cuesta}}{{Caproni} et~al.}{2007}]{Caproni2007}
{Caproni} A.,  {Abraham} Z.,  {Livio} M.,   {Mosquera Cuesta} H.~J.,  2007,
  \mn@doi [\mnras] {10.1111/j.1365-2966.2007.11918.x}, \href
  {http://adsabs.harvard.edu/abs/2007MNRAS.379..135C} {379, 135}

\bibitem[\protect\citeauthoryear{{Chatterjee}, {Liska}, {Tchekhovskoy}  \&
  {Markoff}}{{Chatterjee} et~al.}{2019}]{Chatterjee2019}
{Chatterjee} K.,  {Liska} M.,  {Tchekhovskoy} A.,   {Markoff} S.~B.,  2019,
  MNRAS, submitted (\href{https://arxiv.org/abs/1904.03243}{arXiv:1904.03243}),
  \href {https://ui.adsabs.harvard.edu/abs/2019arXiv190403243C} {}

\bibitem[\protect\citeauthoryear{{Do{\v g}an}, {Nixon}, {King}  \&
  {Pringle}}{{Do{\v g}an} et~al.}{2018}]{Dogan2018}
{Do{\v g}an} S.,  {Nixon} C.~J.,  {King} A.~R.,   {Pringle} J.~E.,  2018,
  \mn@doi [\mnras] {10.1093/mnras/sty155}, \href
  {http://adsabs.harvard.edu/abs/2018MNRAS.476.1519D} {476, 1519}

\bibitem[\protect\citeauthoryear{{Esin}, {McClintock}  \& {Narayan}}{{Esin}
  et~al.}{1997}]{Esin1997}
{Esin} A.~A.,  {McClintock} J.~E.,   {Narayan} R.,  1997, \mn@doi [\apj]
  {10.1086/304829}, \href {http://adsabs.harvard.edu/abs/1997ApJ...489..865E}
  {489, 865}

\bibitem[\protect\citeauthoryear{{Ferreira}, {Petrucci}, {Henri}, {Saug{\'e}}
  \& {Pelletier}}{{Ferreira} et~al.}{2006}]{Fereira2006}
{Ferreira} J.,  {Petrucci} P.~O.,  {Henri} G.,  {Saug{\'e}} L.,   {Pelletier}
  G.,  2006, \mn@doi [\aap] {10.1051/0004-6361:20052689}, \href
  {https://ui.adsabs.harvard.edu/\#abs/2006A&A...447..813F} {447, 813}

\bibitem[\protect\citeauthoryear{{Fiacconi}, {Sijacki}  \&
  {Pringle}}{{Fiacconi} et~al.}{2018}]{Fiaconni2018}
{Fiacconi} D.,  {Sijacki} D.,   {Pringle} J.~E.,  2018, \mn@doi [\mnras]
  {10.1093/mnras/sty893}, \href
  {https://ui.adsabs.harvard.edu/\#abs/2018MNRAS.477.3807F} {477, 3807}

\bibitem[\protect\citeauthoryear{{Fishbone} \& {Moncrief}}{{Fishbone} \&
  {Moncrief}}{1976}]{Fishbone1976}
{Fishbone} L.~G.,  {Moncrief} V.,  1976, \mn@doi [\apj] {10.1086/154565}, \href
  {http://adsabs.harvard.edu/abs/1976ApJ...207..962F} {207, 962}

\bibitem[\protect\citeauthoryear{{Fragile} \& {Blaes}}{{Fragile} \&
  {Blaes}}{2008}]{Fragile2008}
{Fragile} P.~C.,  {Blaes} O.~M.,  2008, \mn@doi [\apj] {10.1086/591936}, \href
  {http://adsabs.harvard.edu/abs/2008ApJ...687..757F} {687, 757}

\bibitem[\protect\citeauthoryear{{Fragile}, {Blaes}, {Anninos}  \&
  {Salmonson}}{{Fragile} et~al.}{2007}]{Fragile2007}
{Fragile} P.~C.,  {Blaes} O.~M.,  {Anninos} P.,   {Salmonson} J.~D.,  2007,
  \mn@doi [\apj] {10.1086/521092}, \href
  {http://adsabs.harvard.edu/abs/2007ApJ...668..417F} {668, 417}

\bibitem[\protect\citeauthoryear{Franchini, Motta  \& Lodato}{Franchini
  et~al.}{2017}]{Franchini2016}
Franchini A.,  Motta S.~E.,   Lodato G.,  2017, \mn@doi [Monthly Notices of the
  Royal Astronomical Society] {10.1093/mnras/stw3363}, 467, 145

\bibitem[\protect\citeauthoryear{{Gammie}, {McKinney}  \& {T{\'o}th}}{{Gammie}
  et~al.}{2003}]{Gammie2003}
{Gammie} C.~F.,  {McKinney} J.~C.,   {T{\'o}th} G.,  2003, \mn@doi [\apj]
  {10.1086/374594}, \href {http://adsabs.harvard.edu/abs/2003ApJ...589..444G}
  {589, 444}

\bibitem[\protect\citeauthoryear{{Gardiner} \& {Stone}}{{Gardiner} \&
  {Stone}}{2005}]{Gardiner2005}
{Gardiner} T.~A.,  {Stone} J.~M.,  2005, \mn@doi [Journal of Computational
  Physics] {10.1016/j.jcp.2004.11.016}, \href
  {http://adsabs.harvard.edu/abs/2005JCoPh.205..509G} {205, 509}

\bibitem[\protect\citeauthoryear{{Greene}, {Bailyn}  \& {Orosz}}{{Greene}
  et~al.}{2001}]{Greene2001}
{Greene} J.,  {Bailyn} C.~D.,   {Orosz} J.~A.,  2001, \mn@doi [\apj]
  {10.1086/321411}, \href {http://adsabs.harvard.edu/abs/2001ApJ...554.1290G}
  {554, 1290}

\bibitem[\protect\citeauthoryear{{Hawley} \& {Krolik}}{{Hawley} \&
  {Krolik}}{2019}]{Hawley2019}
{Hawley} J.~F.,  {Krolik} J.~H.,  2019, \mn@doi [\apj]
  {10.3847/1538-4357/ab1f6e}, \href
  {https://ui.adsabs.harvard.edu/abs/2019ApJ...878..149H} {878, 149}

\bibitem[\protect\citeauthoryear{{Hjellming} \& {Rupen}}{{Hjellming} \&
  {Rupen}}{1995}]{Hjelming1995}
{Hjellming} R.~M.,  {Rupen} M.~P.,  1995, \mn@doi [\nat] {10.1038/375464a0},
  \href {http://adsabs.harvard.edu/abs/1995Natur.375..464H} {375, 464}

\bibitem[\protect\citeauthoryear{{Ingram}, {Done}  \& {Fragile}}{{Ingram}
  et~al.}{2009}]{Ingram2009}
{Ingram} A.,  {Done} C.,   {Fragile} P.~C.,  2009, \mn@doi [\mnras]
  {10.1111/j.1745-3933.2009.00693.x}, \href
  {http://adsabs.harvard.edu/abs/2009MNRAS.397L.101I} {397, L101}

\bibitem[\protect\citeauthoryear{{Ingram}, {van der Klis}, {Middleton}, {Done},
  {Altamirano}, {Heil}, {Uttley}  \& {Axelsson}}{{Ingram}
  et~al.}{2016}]{Ingram2016}
{Ingram} A.,  {van der Klis} M.,  {Middleton} M.,  {Done} C.,  {Altamirano} D.,
   {Heil} L.,  {Uttley} P.,   {Axelsson} M.,  2016, \mn@doi [\mnras]
  {10.1093/mnras/stw1245}, \href
  {http://adsabs.harvard.edu/abs/2016MNRAS.461.1967I} {461, 1967}

\bibitem[\protect\citeauthoryear{{Ivanov} \& {Illarionov}}{{Ivanov} \&
  {Illarionov}}{1997}]{Ivanov1997}
{Ivanov} P.~B.,  {Illarionov} A.~F.,  1997, \mn@doi [\mnras]
  {10.1093/mnras/285.2.394}, \href
  {http://adsabs.harvard.edu/abs/1997MNRAS.285..394I} {285, 394}

\bibitem[\protect\citeauthoryear{{Jiang}, {Stone}  \& {Davis}}{{Jiang}
  et~al.}{2017}]{Jiang2017}
{Jiang} Y.-F.,  {Stone} J.,   {Davis} S.~W.,  2017, ApJ, submitted
  (\href{https://arxiv.org/abs/1709.02845}{arXiv:1709.02845}), \href
  {http://adsabs.harvard.edu/abs/2017arXiv170902845J} {}

\bibitem[\protect\citeauthoryear{{Kalamkar}, {Casella}, {Uttley}, {O'Brien},
  {Russell}, {Maccarone}, {van der Klis}  \& {Vincentelli}}{{Kalamkar}
  et~al.}{2016}]{Kalamkar2016}
{Kalamkar} M.,  {Casella} P.,  {Uttley} P.,  {O'Brien} K.,  {Russell} D.,
  {Maccarone} T.,  {van der Klis} M.,   {Vincentelli} F.,  2016, \mn@doi
  [\mnras] {10.1093/mnras/stw1211}, \href
  {http://adsabs.harvard.edu/abs/2016MNRAS.460.3284K} {460, 3284}

\bibitem[\protect\citeauthoryear{{King} \& {Pringle}}{{King} \&
  {Pringle}}{2006}]{King2006}
{King} A.~R.,  {Pringle} J.~E.,  2006, \mn@doi [\mnras]
  {10.1111/j.1745-3933.2006.00249.x}, \href
  {https://ui.adsabs.harvard.edu/abs/2006MNRAS.373L..90K} {373, L90}

\bibitem[\protect\citeauthoryear{{King}, {Lubow}, {Ogilvie}  \&
  {Pringle}}{{King} et~al.}{2005}]{King2005}
{King} A.~R.,  {Lubow} S.~H.,  {Ogilvie} G.~I.,   {Pringle} J.~E.,  2005,
  \mn@doi [\mnras] {10.1111/j.1365-2966.2005.09378.x}, \href
  {http://adsabs.harvard.edu/abs/2005MNRAS.363...49K} {363, 49}

\bibitem[\protect\citeauthoryear{{Kumar} \& {Pringle}}{{Kumar} \&
  {Pringle}}{1985}]{Kumar1985}
{Kumar} S.,  {Pringle} J.~E.,  1985, \mn@doi [\mnras]
  {10.1093/mnras/213.3.435}, \href
  {http://adsabs.harvard.edu/abs/1985MNRAS.213..435K} {213, 435}

\bibitem[\protect\citeauthoryear{{Lense} \& {Thirring}}{{Lense} \&
  {Thirring}}{1918}]{LT1918}
{Lense} J.,  {Thirring} H.,  1918, Physikalische Zeitschrift, \href
  {http://adsabs.harvard.edu/abs/1918PhyZ...19..156L} {19}

\bibitem[\protect\citeauthoryear{{Liska}, {Tchekhovskoy}  \&
  {Quataert}}{{Liska} et~al.}{2018a}]{Liska2018B}
{Liska} M.~T.~P.,  {Tchekhovskoy} A.,   {Quataert} E.,  2018a, arXiv e-prints,
  \href {http://adsabs.harvard.edu/abs/2018arXiv180904608L} {}

\bibitem[\protect\citeauthoryear{{Liska}, {Hesp}, {Tchekhovskoy}, {Ingram},
  {van der Klis}  \& {Markoff}}{{Liska} et~al.}{2018b}]{Liska2018A}
{Liska} M.,  {Hesp} C.,  {Tchekhovskoy} A.,  {Ingram} A.,  {van der Klis} M.,
  {Markoff} S.,  2018b, \mn@doi [\mnras] {10.1093/mnrasl/slx174}, \href
  {http://adsabs.harvard.edu/abs/2018MNRAS.474L..81L} {474, L81}

\bibitem[\protect\citeauthoryear{{Liska}, {Hesp}, {Tchekhovskoy}, {Ingram},
  {van der Klis}  \& {Markoff}}{{Liska} et~al.}{2019a}]{Liska2019A}
{Liska} M.,  {Hesp} C.,  {Tchekhovskoy} A.,  {Ingram} A.,  {van der Klis} M.,
  {Markoff} S.~B.,  2019a, MNRAS, submitted
  (\href{https://arxiv.org/abs/1901.05970}{arXiv:1901.05970}), \href
  {https://ui.adsabs.harvard.edu/\#abs/2019arXiv190105970L} {}

\bibitem[\protect\citeauthoryear{{Liska}, {Tchekhovskoy}, {Ingram}  \& {van der
  Klis}}{{Liska} et~al.}{2019b}]{Liska2018C}
{Liska} M.,  {Tchekhovskoy} A.,  {Ingram} A.,   {van der Klis} M.,  2019b,
  MNRAS, in press (\href{https://arxiv.org/abs/1810.00883}{arXiv:1810.00883}),
  \href {https://ui.adsabs.harvard.edu/#abs/2018arXiv181000883L} {}

\bibitem[\protect\citeauthoryear{Liska et~al.,}{Liska et~al.}{2019c}]{hamr}
Liska M.,  et~al., 2019c, MNRAS, submitted
  (\href{https://arxiv.org/abs/1912.10192}{arXiv:1912.10192})

\bibitem[\protect\citeauthoryear{{Lodato} \& {Price}}{{Lodato} \&
  {Price}}{2010}]{Lodato2010}
{Lodato} G.,  {Price} D.~J.,  2010, \mn@doi [\mnras]
  {10.1111/j.1365-2966.2010.16526.x}, \href
  {http://adsabs.harvard.edu/abs/2010MNRAS.405.1212L} {405, 1212}

\bibitem[\protect\citeauthoryear{Lodato \& Pringle}{Lodato \&
  Pringle}{2006}]{Lodato06}
Lodato G.,  Pringle J.~E.,  2006, \mn@doi [Monthly Notices of the Royal
  Astronomical Society] {10.1111/j.1365-2966.2006.10194.x}, 368, 1196

\bibitem[\protect\citeauthoryear{{Lubow}, {Ogilvie}  \& {Pringle}}{{Lubow}
  et~al.}{2002}]{Lubow2002}
{Lubow} S.~H.,  {Ogilvie} G.~I.,   {Pringle} J.~E.,  2002, \mn@doi [\mnras]
  {10.1046/j.1365-8711.2002.05949.x}, \href
  {http://adsabs.harvard.edu/abs/2002MNRAS.337..706L} {337, 706}

\bibitem[\protect\citeauthoryear{{Marcel} et~al.,}{{Marcel}
  et~al.}{2018a}]{Marcel2018A}
{Marcel} G.,  et~al., 2018a, \mn@doi [\aap] {10.1051/0004-6361/201732069},
  \href {http://adsabs.harvard.edu/abs/2018A%26A...615A..57M} {615, A57}

\bibitem[\protect\citeauthoryear{{Marcel} et~al.,}{{Marcel}
  et~al.}{2018b}]{Marcel2018B}
{Marcel} G.,  et~al., 2018b, \mn@doi [\aap] {10.1051/0004-6361/201833124},
  \href {http://adsabs.harvard.edu/abs/2018A%26A...617A..46M} {617, A46}

\bibitem[\protect\citeauthoryear{{McClintock} \& {Remillard}}{{McClintock} \&
  {Remillard}}{2006}]{McClintock2006}
{McClintock} J.~E.,  {Remillard} R.~A.,  2006, {Black hole binaries}.
pp 157--213

\bibitem[\protect\citeauthoryear{{McClintock}, {Narayan}  \&
  {Steiner}}{{McClintock} et~al.}{2014}]{Mcclintock2013}
{McClintock} J.~E.,  {Narayan} R.,   {Steiner} J.~F.,  2014, \mn@doi [\ssr]
  {10.1007/s11214-013-0003-9}, \href
  {http://adsabs.harvard.edu/abs/2014SSRv..183..295M} {183, 295}

\bibitem[\protect\citeauthoryear{{McKinney}, {Tchekhovskoy}  \&
  {Blandford}}{{McKinney} et~al.}{2012}]{Mckinney2012}
{McKinney} J.~C.,  {Tchekhovskoy} A.,   {Blandford} R.~D.,  2012, \mn@doi
  [\mnras] {10.1111/j.1365-2966.2012.21074.x}, \href
  {http://adsabs.harvard.edu/abs/2012MNRAS.423.3083M} {423, 3083}

\bibitem[\protect\citeauthoryear{{Morales Teixeira}, {Fragile}, {Zhuravlev}  \&
  {Ivanov}}{{Morales Teixeira} et~al.}{2014}]{Texeira2014}
{Morales Teixeira} D.,  {Fragile} P.~C.,  {Zhuravlev} V.~V.,   {Ivanov} P.~B.,
  2014, \mn@doi [\apj] {10.1088/0004-637X/796/2/103}, \href
  {http://adsabs.harvard.edu/abs/2014ApJ...796..103M} {796, 103}

\bibitem[\protect\citeauthoryear{Motta, Franchini, Lodato  \&
  Mastroserio}{Motta et~al.}{2018}]{Motta2017}
Motta S.~E.,  Franchini A.,  Lodato G.,   Mastroserio G.,  2018, \mn@doi
  [Monthly Notices of the Royal Astronomical Society] {10.1093/mnras/stx2358},
  473, 431

\bibitem[\protect\citeauthoryear{{Narayan} \& {Yi}}{{Narayan} \&
  {Yi}}{1994}]{Narayan1994}
{Narayan} R.,  {Yi} I.,  1994, \mn@doi [\apjl] {10.1086/187381}, \href
  {http://adsabs.harvard.edu/abs/1994ApJ...428L..13N} {428, L13}

\bibitem[\protect\citeauthoryear{{Natarajan} \& {Pringle}}{{Natarajan} \&
  {Pringle}}{1998}]{Natarajan1998}
{Natarajan} P.,  {Pringle} J.~E.,  1998, \mn@doi [\apjl] {10.1086/311658},
  \href {http://adsabs.harvard.edu/abs/1998ApJ...506L..97N} {506, L97}

\bibitem[\protect\citeauthoryear{{Nealon}, {Price}  \& {Nixon}}{{Nealon}
  et~al.}{2015}]{Nealon2015}
{Nealon} R.,  {Price} D.~J.,   {Nixon} C.~J.,  2015, \mn@doi [\mnras]
  {10.1093/mnras/stv014}, \href
  {http://adsabs.harvard.edu/abs/2015MNRAS.448.1526N} {448, 1526}

\bibitem[\protect\citeauthoryear{{Nelson} \& {Papaloizou}}{{Nelson} \&
  {Papaloizou}}{2000}]{Nelson2000}
{Nelson} R.~P.,  {Papaloizou} J.~C.~B.,  2000, \mn@doi [\mnras]
  {10.1046/j.1365-8711.2000.03478.x}, \href
  {http://adsabs.harvard.edu/abs/2000MNRAS.315..570N} {315, 570}

\bibitem[\protect\citeauthoryear{{Nixon} \& {King}}{{Nixon} \&
  {King}}{2012}]{Nixon2012A}
{Nixon} C.~J.,  {King} A.~R.,  2012, \mn@doi [\mnras]
  {10.1111/j.1365-2966.2011.20377.x}, \href
  {http://adsabs.harvard.edu/abs/2012MNRAS.421.1201N} {421, 1201}

\bibitem[\protect\citeauthoryear{{Nixon}, {King}  \& {Price}}{{Nixon}
  et~al.}{2012a}]{Nixon2012counter}
{Nixon} C.~J.,  {King} A.~R.,   {Price} D.~J.,  2012a, \mn@doi [\mnras]
  {10.1111/j.1365-2966.2012.20814.x}, \href
  {https://ui.adsabs.harvard.edu/\#abs/2012MNRAS.422.2547N} {422, 2547}

\bibitem[\protect\citeauthoryear{{Nixon}, {King}, {Price}  \& {Frank}}{{Nixon}
  et~al.}{2012b}]{Nixon2012B}
{Nixon} C.,  {King} A.,  {Price} D.,   {Frank} J.,  2012b, \mn@doi [\apjl]
  {10.1088/2041-8205/757/2/L24}, \href
  {http://adsabs.harvard.edu/abs/2012ApJ...757L..24N} {757, L24}

\bibitem[\protect\citeauthoryear{{Noble}, {Krolik}  \& {Hawley}}{{Noble}
  et~al.}{2009}]{Noble2009}
{Noble} S.~C.,  {Krolik} J.~H.,   {Hawley} J.~F.,  2009, \mn@doi [\apj]
  {10.1088/0004-637X/692/1/411}, \href
  {http://adsabs.harvard.edu/abs/2009ApJ...692..411N} {692, 411}

\bibitem[\protect\citeauthoryear{{Ogilvie}}{{Ogilvie}}{1999}]{Ogilvie1999}
{Ogilvie} G.~I.,  1999, \mn@doi [\mnras] {10.1046/j.1365-8711.1999.02340.x},
  \href {http://adsabs.harvard.edu/abs/1999MNRAS.304..557O} {304, 557}

\bibitem[\protect\citeauthoryear{{Orosz} et~al.,}{{Orosz}
  et~al.}{2001}]{Orosz2001}
{Orosz} J.~A.,  et~al., 2001, \mn@doi [\apj] {10.1086/321442}, \href
  {https://ui.adsabs.harvard.edu/abs/2001ApJ...555..489O} {555, 489}

\bibitem[\protect\citeauthoryear{{Papaloizou} \& {Lin}}{{Papaloizou} \&
  {Lin}}{1995}]{papaloizou1995}
{Papaloizou} J.~C.~B.,  {Lin} D.~N.~C.,  1995, \mn@doi [\araa]
  {10.1146/annurev.aa.33.090195.002445}, \href
  {http://adsabs.harvard.edu/abs/1995ARA%26A..33..505P} {33, 505}

\bibitem[\protect\citeauthoryear{{Papaloizou} \& {Pringle}}{{Papaloizou} \&
  {Pringle}}{1983}]{Papaloizou1983}
{Papaloizou} J.~C.~B.,  {Pringle} J.~E.,  1983, \mn@doi [\mnras]
  {10.1093/mnras/202.4.1181}, \href
  {http://adsabs.harvard.edu/abs/1983MNRAS.202.1181P} {202, 1181}

\bibitem[\protect\citeauthoryear{{Penna}, {McKinney}, {Narayan},
  {Tchekhovskoy}, {Shafee}  \& {McClintock}}{{Penna} et~al.}{2010}]{Penna2010}
{Penna} R.~F.,  {McKinney} J.~C.,  {Narayan} R.,  {Tchekhovskoy} A.,  {Shafee}
  R.,   {McClintock} J.~E.,  2010, \mn@doi [\mnras]
  {10.1111/j.1365-2966.2010.17170.x}, \href
  {http://adsabs.harvard.edu/abs/2010MNRAS.408..752P} {408, 752}

\bibitem[\protect\citeauthoryear{{Remillard} \& {McClintock}}{{Remillard} \&
  {McClintock}}{2006}]{Remillard2006}
{Remillard} R.~A.,  {McClintock} J.~E.,  2006, \mn@doi [\araa]
  {10.1146/annurev.astro.44.051905.092532}, \href
  {http://adsabs.harvard.edu/abs/2006ARA%26A..44...49R} {44, 49}

\bibitem[\protect\citeauthoryear{{Reynolds} \& {Fabian}}{{Reynolds} \&
  {Fabian}}{2008}]{Reynolds2008}
{Reynolds} C.~S.,  {Fabian} A.~C.,  2008, \mn@doi [\apj] {10.1086/527344},
  \href {http://adsabs.harvard.edu/abs/2008ApJ...675.1048R} {675, 1048}

\bibitem[\protect\citeauthoryear{{Shakura} \& {Sunyaev}}{{Shakura} \&
  {Sunyaev}}{1973}]{ss73}
{Shakura} N.~I.,  {Sunyaev} R.~A.,  1973, \aap, \href
  {http://adsabs.harvard.edu/abs/1973A%26A....24..337S} {24, 337}

\bibitem[\protect\citeauthoryear{{Shiokawa}, {Dolence}, {Gammie}  \&
  {Noble}}{{Shiokawa} et~al.}{2012}]{Hawley2011}
{Shiokawa} H.,  {Dolence} J.~C.,  {Gammie} C.~F.,   {Noble} S.~C.,  2012,
  \mn@doi [\apj] {10.1088/0004-637X/744/2/187}, \href
  {https://ui.adsabs.harvard.edu/\#abs/2012ApJ...744..187S} {744, 187}

\bibitem[\protect\citeauthoryear{{Sorathia}, {Reynolds}  \&
  {Armitage}}{{Sorathia} et~al.}{2010}]{Sorathia2010}
{Sorathia} K.~A.,  {Reynolds} C.~S.,   {Armitage} P.~J.,  2010, \mn@doi [\apj]
  {10.1088/0004-637X/712/2/1241}, \href
  {http://adsabs.harvard.edu/abs/2010ApJ...712.1241S} {712, 1241}

\bibitem[\protect\citeauthoryear{{Stella} \& {Vietri}}{{Stella} \&
  {Vietri}}{1998}]{Stella1998}
{Stella} L.,  {Vietri} M.,  1998, \mn@doi [\apjl] {10.1086/311075}, \href
  {http://adsabs.harvard.edu/abs/1998ApJ...492L..59S} {492, L59}

\bibitem[\protect\citeauthoryear{{Stevens} \& {Uttley}}{{Stevens} \&
  {Uttley}}{2016}]{Stevens2016}
{Stevens} A.~L.,  {Uttley} P.,  2016, \mn@doi [\mnras] {10.1093/mnras/stw1093},
  \href {https://ui.adsabs.harvard.edu/\#abs/2016MNRAS.460.2796S} {460, 2796}

\bibitem[\protect\citeauthoryear{{Tchekhovskoy}, {Narayan}  \&
  {McKinney}}{{Tchekhovskoy} et~al.}{2011}]{Tchekhovskoy2011}
{Tchekhovskoy} A.,  {Narayan} R.,   {McKinney} J.~C.,  2011, \mn@doi [\mnras]
  {10.1111/j.1745-3933.2011.01147.x}, \href
  {http://adsabs.harvard.edu/abs/2011MNRAS.418L..79T} {418, L79}

\bibitem[\protect\citeauthoryear{{White}, {Quataert}  \& {Blaes}}{{White}
  et~al.}{2019}]{White2019A}
{White} C.~J.,  {Quataert} E.,   {Blaes} O.,  2019, ApJ, submitted
  (\href{https://arxiv.org/abs/1902.09662}{arXiv:1902.09662}), \href
  {https://ui.adsabs.harvard.edu/abs/2019arXiv190209662W} {}

\bibitem[\protect\citeauthoryear{van~der Klis}{van~der Klis}{1989}]{Klis1989}
van~der Klis M.,  1989, \mn@doi [Annual Review of Astronomy and Astrophysics]
  {10.1146/annurev.aa.27.090189.002505}, 27, 517

\makeatother
\end{thebibliography}
\bibliographystyle{mnras}
\label{lastpage}
\end{document}